\documentclass[twocolumn]{aastex62}
\pdfoutput=1
\usepackage{natbib}
\def\be{\begin{equation}}
\def\ee{\end{equation}}

\def\ujybm{~$\mu$Jy~bm$^{-1}$}

\def\um{~$\mu$m}
\def\ghz{~GHz}

\def\HII{\ion{H}{2}}

\def\alma{{\it ALMA}}
\def\hst{{\it HST}}
\def\galex{{\it GALEX}}

\def\jwst{{\it JWST}}
\def\spitzer{{\it Spitzer}}

\begin{document}
\title{An ALMA-HST Study of Millimeter Dust Emission and Star Clusters}
\shorttitle{ALMA-HST Dust Emission and Star Clusters}
\shortauthors{Turner et al.}

\author[0000-0003-2261-5746]{J.~A. Turner}
\affiliation{Department of Physics \& Astronomy, University of Wyoming, Laramie WY, USA; jturne19@uwyo.edu}

\author{D.~A. Dale}
\affiliation{Department of Physics \& Astronomy, University of Wyoming, Laramie WY, USA}

\author{A. Adamo}
\affiliation{Dept. of Astronomy, The Oskar Klein Centre, Stockholm University, Stockholm, Sweden}

\author{D. Calzetti}
\affiliation{Department of Astronomy, University of Massachusetts, Amherst MA, USA}

\author{K. Grasha}
\affiliation{Research School of Astronomy and Astrophysics, Australian National University, Canberra, Australia}

\author{E. K. Grebel}
\affiliation{Astronomisches Rechen-Institut, Zentrum f{\"u}r Astronomie der Universit{\"a}t Heidelberg, Heidelberg, Germany}

\author{K. E. Johnson}
\affiliation{Department of Astronomy, University of Virginia, Charlottesville VA, USA}

\author{J. C. Lee}
\affiliation{California Institute of Technology/IPAC, Pasadena CA, USA}

\author{L. J. Smith}
\affiliation{Space Telescope Science Institute and European Space Agency, 3700 San Martin Drive, Baltimore, MD 21218}

\author{I. Yoon}
\affiliation{National Radio Astronomy Observatory, Charlottesville VA,  USA}
\begin{abstract}
We present results from a joint {\it ALMA-HST} study of the nearby spiral galaxy NGC~628. We combine the \hst\ LEGUS database of over 1000 stellar clusters in NGC~628 with \alma\ Cycle 4 millimeter/submillimeter observations of the cold dust continuum that span $\sim$15~kpc$^2$ including the nuclear region and western portions of the galaxy's disk. The resolution---1\farcs1 or approximately 50 pc at the distance of NGC~628---allows us to constrain the spatial variations in the slope of the millimeter dust continuum as a function of the ages and masses of the nearby stellar clusters. Our results indicate an excess of dust emission in the millimeter assuming a typical cold dust model for a normal star-forming galaxy, but little correlation of the dust continuum slope with stellar cluster age or mass. For the depth and spatial coverage of these observations, we cannot substantiate the millimeter/submillimeter excess arising from the processing of dust grains by the local interstellar radiation field. We detect a bright unknown source in NGC~628 in \alma\ bands 4 and 7 with no counterparts at other wavelengths from ancillary data. We speculate this is possibly a dust obscured supernova.

\end{abstract}


\section{Introduction}
\label{sec:intro}

Many galaxies are factories of current star formation that deplete the gas of the interstellar medium (ISM). Dense regions of gas condense into stars and star clusters which exhaust the galaxy of gas; in some instances, the gas is replenished through mergers or by the infall of external gas. Star formation, in turn, affects the evolution of galaxies by returning metals, energy, and momentum into the ISM and intergalactic medium. The cycling of gas into and out of galaxies regulates galaxy growth and its turbulence regulates star formation. In order to understand galaxy evolution, it is crucial to understand the physical processes that determine the evolution. Emission from stars and star clusters, along with the dust and gas of the ISM, all contribute to the shape of a galaxy's spectral energy distribution (SED). By studying spatially-resolved SEDs, we can begin to understand the small and large-scale properties of galaxies and the astrophysical processes underlying their evolution.

One unresolved question of using SEDs to study the star-forming environments of galaxies is an observed excess at millimeter/submillimeter wavelengths. This excess has been primarily seen in low-metallicity star-forming galaxies like the Magellanic Clouds \citep[i.e.,][]{Galliano2005,Galametz2011,Planck2011,Gordon2014,Izotov2014,Hermelo2016,Dale2017}. Dust emission in the far-infrared regime of the SED is often described as a modified blackbody, $S_{\nu} \propto B_{\nu}(T_{\text{dust}})\nu^{\beta}$, where $\beta$ is a measured effective grain emissivity parameter that empirically ranges between $\sim$\,0.8 and 2.5. This emissivity has been seen to vary significantly not only between galaxies \citep{Galametz2012} but also within galaxies \citep{Kirkpatrick2014}. In some cases, a modified blackbody extrapolated from far-infrared photometry cannot fully account for the emission beyond $\sim$\,500\um, hence the so-called `millimeter/submillimeter excess' \citep{Galliano2003,Bot2010,Planck2011}. Proposed causes of this excess include very cold dust at T$_{\text{dust}}<$~7~K \citep{Galliano2005}; fluctuations in the cosmic microwave background radiation as found in the LMC \citep{Planck2011}; thermal free-free emission of ionized gas \citep{Izotov2014}; non-thermal emission from polycyclic aromatic hydrocarbons or nanoparticles \citep{Lisenfeld2002,Meny2007,Coupeaud2011,Hensley2017}; and dust emissivity variations can be caused by processing of the dust that results in a flatter $\beta$ which we explore here \citep{Gordon2014,Hermelo2016}. 

\cite{Kirkpatrick2014} study the spatial variations of $\beta$ in the KINGFISH (Key Insights on Nearby Galaxies: A Far-Infrared Survey with {\it Herschel}) sample of nearby star-forming galaxies and find a radial dependence for $\beta$ which flattens towards the outskirts of the galaxies. The authors test the possibility that $\beta$ is modified by different heating sources, specifically young versus old stellar populations. The two stellar populations both contribute to the interstellar radiation field (ISRF) with the younger stars providing comparatively more ionizing photons than the less energetic photons from the old population. They find a correlation between $\beta$ and the heating from the old stellar population as quantified by the ratio of luminosities at 3.6 and 500\um. This correlation is interpreted to mean the submillimeter slope is flattened due to inefficient heating of the dust grains by the old stellar population. The inefficient heating could allow a cold dust emission component to exist at wavelengths beyond that of the dust emission peak, a scenario that will manifest as a shallower $\beta$. 

The {\it Hubble Space Telescope} treasury program Legacy ExtraGalactic UV Survey (LEGUS) was designed to provide a novel catalog of star clusters in 50 nearby galaxies and their properties \citep[see][]{Calzetti2015}. Given the \hst\ resolution and proximity of the LEGUS targets, the galaxies have been resolved into their main stellar components: stars, star clusters, and associations. The LEGUS dataset provides star clusters' measured properties including age, mass, and color excess for a broad range of cluster populations across the LEGUS targets \citep[see][]{Adamo2017}. In the LEGUS cluster catalog, the average uncertainty for both a cluster's age and mass is 0.1 dex.

\begin{figure*}
\epsscale{0.55}
\plotone{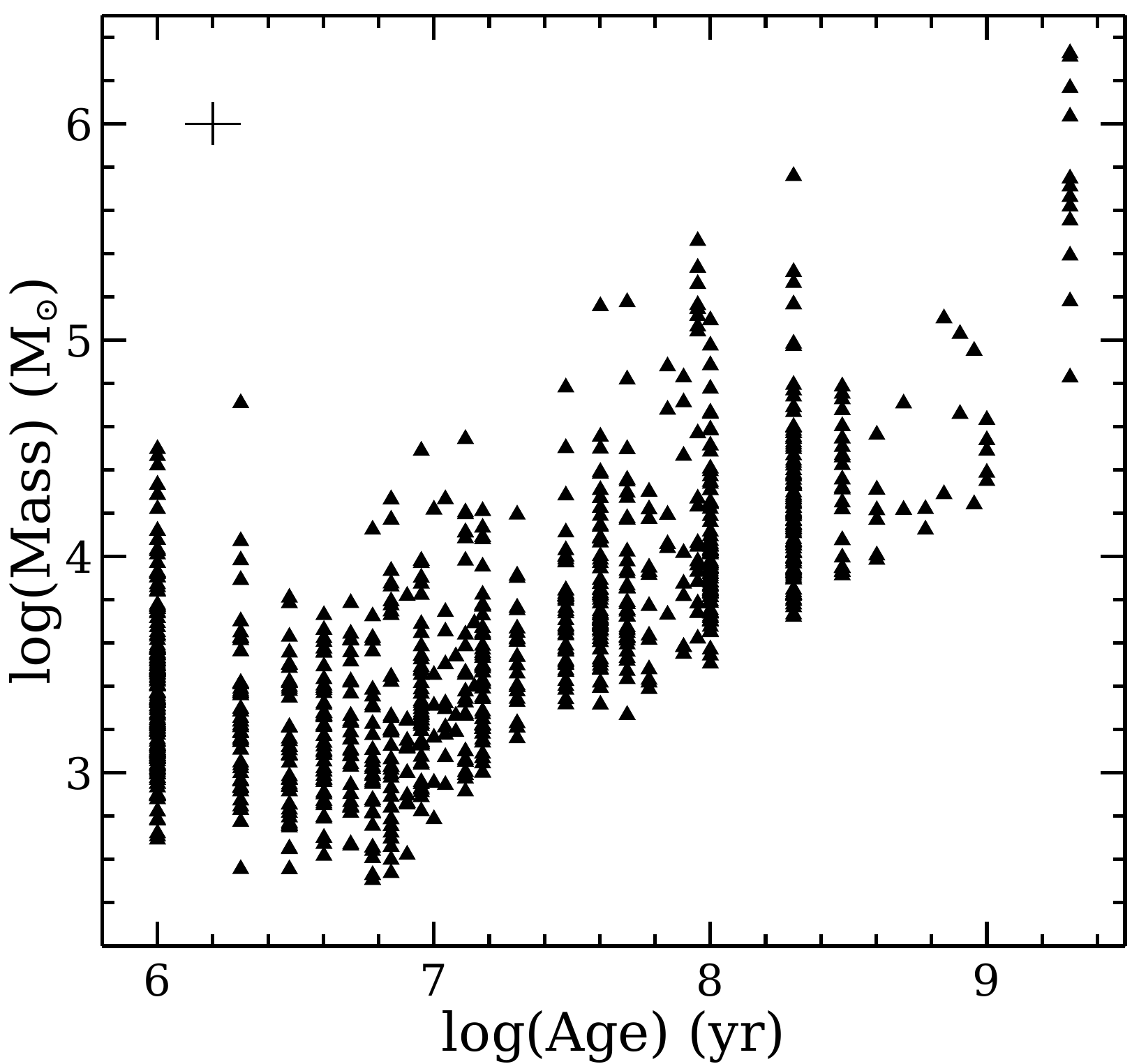}
\caption{Distribution of the ages and masses of the LEGUS classified star clusters in the central field of NGC~628. A typical error bar is given in the top left corner. Data from \cite{Grasha2015,Adamo2017}.}
\label{fig:cluster}
\end{figure*}

\cite{Grasha2019} make use of the rich LEGUS dataset by exploring the spatial relation between star clusters and giant molecular clouds (GMCs) in the spiral galaxy NGC~5194 (M51). The authors find a spatial correlation between young ($\leq$10~Myr) star clusters and the GMCs which gives a time-scale for disassociation of the star clusters from the GMCs of $\sim$\,4--6~Myr. A similar study from \cite{Matthews2018} compares \hst\ and \alma\ observations and determines that by 10$^{6.7}$ years, some star clusters will have lost all of their molecular material from which they were formed. In this study, we look to link the star clusters, not with their associated gas, but with the dust leftover from the cluster formation.

In this work, we combine the LEGUS star cluster catalog with dust continuum observations from the Atacama Large Millimeter/Submillimeter Array (\alma) of the nearby spiral galaxy NGC~628 (M74). NGC~628 is nearly face-on ($i \sim 25$\degr), located at a distance of 9.9~Mpc \citep{Calzetti2015}, and provides an excellent testbed for these unprecedented \alma\ observations with the wealth of LEGUS star clusters and ancillary data available. Figure~\ref{fig:cluster} shows the distribution of the star cluster ages and masses provided by LEGUS in NGC~628. We make use of the LEGUS cluster catalog with the averaged aperture corrections, Milky Way extinction, and Padova-AGB stellar evolution track\footnote{The star cluster catalog is available at\dataset[legus.stsci.edu]{https://legus.stsci.edu}.}. By combining the high resolution \alma\ dust continuum maps with the star cluster data from \hst, we can begin to study how the local ISRF generated by the star clusters may be affecting the dust emissivity.

In Section~\ref{sec:obs}, we describe the new observations from ALMA.  In Section~\ref{sec:analysis}, we review the analysis carried out, and in Section~\ref{sec:results}, we present the essential results and discuss their implications.  Finally, in Section~\ref{sec:sum}, we summarize our findings.

\section{\alma\ Observations}
\label{sec:obs}

The \alma\ Cycle~4 observations in band~7 (343\ghz, 0.87~mm) and band~4 (145\ghz, 2.1~mm) were carried out in 2016--2017 (ID = 2016.1.01435.S3, PI = D. Dale). Our band~7 observations consisted of 137 pointings with 43 12-meter antennas with a baseline ranging from 15.1~m to 2.6~km. The phase and amplitude were calibrated using sources J0006-0623 and J0121+1149.  Our band~4 observations consisted of 23 pointings with 44 12-meter antennas with a baseline ranging from 18.6~m to 1.1~km. The phase and amplitude were calibrated using sources J0006-0623, J0139+1753, and J0238+1636. 

The data were reduced using the Common Astronomy Software Applications (CASA) package version 5.4.0-68 \citep{casa2007}. Images were constructed and cleaned using the CASA task {\tt TCLEAN} in the mosaic imaging mode. A Briggs weighting is used with the ``robust'' parameter set to 2.0 in order to increase the sensitivity to extended emission at the cost of angular resolution. A $u$-$v$ taper of 1\farcs2 and 1\farcs0 is applied to the band~7 and band~4 images, respectively. This gives a restoring beam size of 1\farcs12 $\times$ 1\farcs04 with position angle 39.73$\degr$ for band~7 with a maximum recoverable scale of 7\farcs2. For band~4, the restoring beam is 1\farcs12$\times$1\farcs08 with position angle 29.63$\degr$ and a maximum recoverable scale 13\farcs8. The angular resolution of 1\farcs1 corresponds to a spatial resolution of about 50~pc assuming a distance of 9.9~Mpc to NGC~628 \citep{Calzetti2015}. At these resolutions, the 1$\sigma$ sensitivity is 225\ujybm\ for band~7 and 31.7\ujybm\ for band~4. Continuum maps were generated using all four spectral windows (SPWs 17, 19, 21, 23) in each band which were then corrected for the primary beam attenuation and used for the analysis in this paper. The sensitivity was measured on the images before the primary beam correction. 

\begin{figure*}
\epsscale{0.8}
\plotone{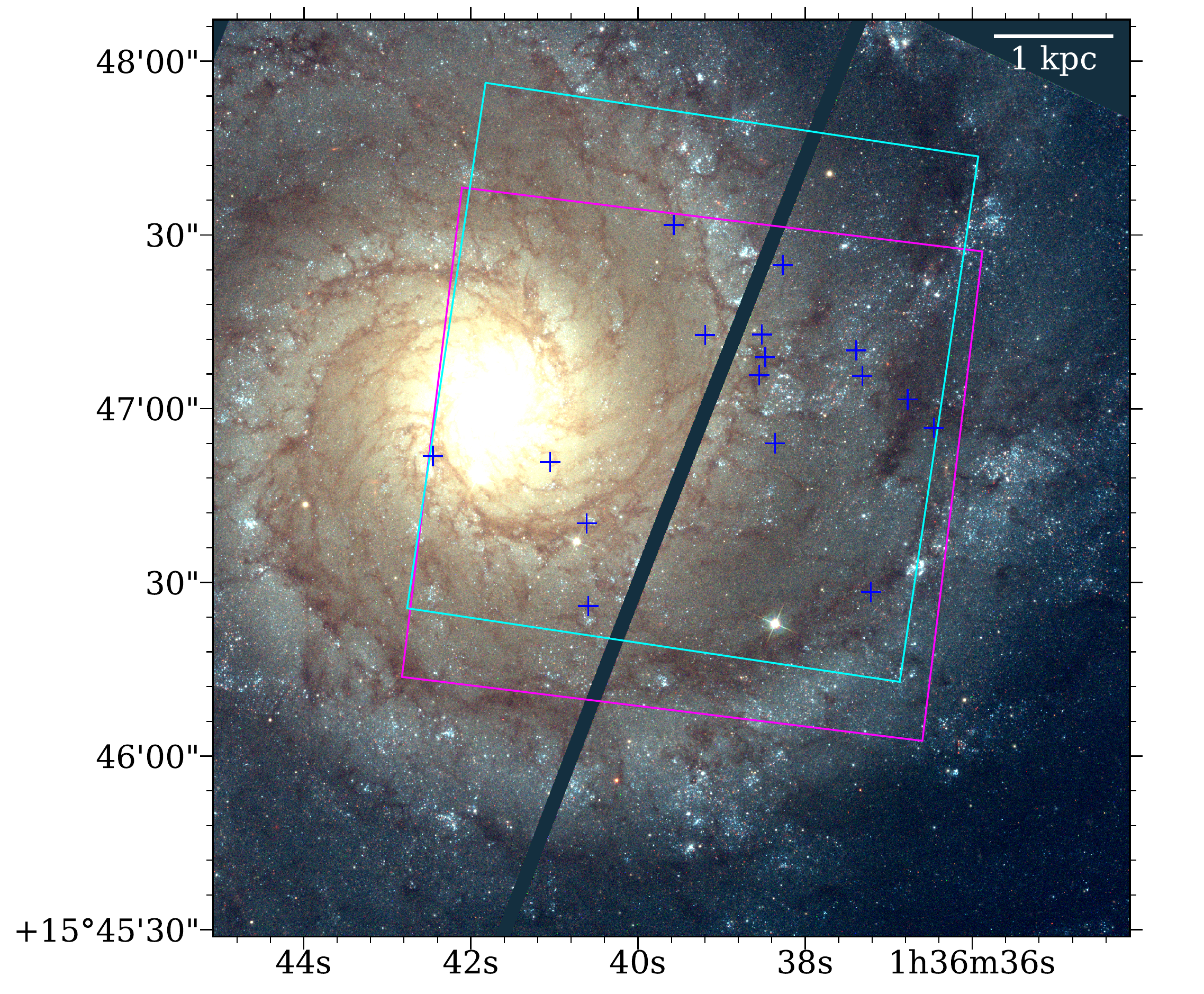}
\caption{A 3-color image of NGC~628 using \hst\ F435W, F555W, and F814W observations from LEGUS. Overlaid are the 80\% primary-beam coverage footprints from the \alma\ band 7 (cyan) and band 4 (magenta) maps. The center locations of the 16 millimeter sources as shown in Table~1 are marked by the blue crosses.}
\label{fig:hst}
\end{figure*}

In both bands, a `footprint' is defined as $\geq$80\% of the primary-beam coverage. Figure~\ref{fig:hst} shows rectangular approximations for the two footprints overlaid on a 3-color image using HST LEGUS observations. For the analysis carried out here, we focus on the region where the footprints overlap which corresponds to about 1.8~arcmin$^{2}$ or 15.1~kpc$^{2}$. This area encompasses the center of the galaxy out to 0.27~R$_{25}$ ($\sim$\,4~kpc), where $R_{25}$ is the traditional 25~${\rm mag}~{\rm arcsec}^{-2}$ isophotal radius.

\section{Analysis}
\label{sec:analysis}

\begin{figure*}
\plottwo{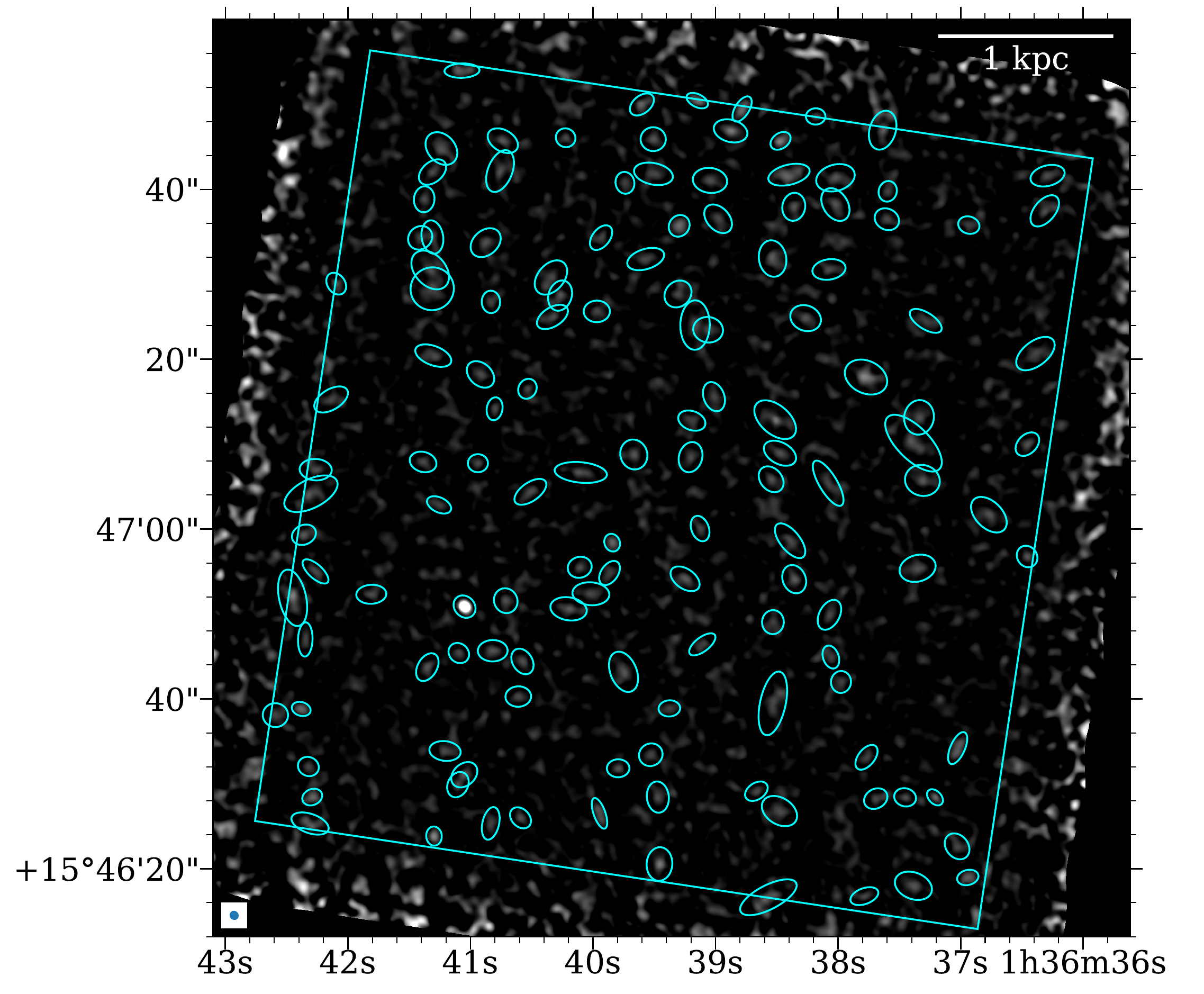}{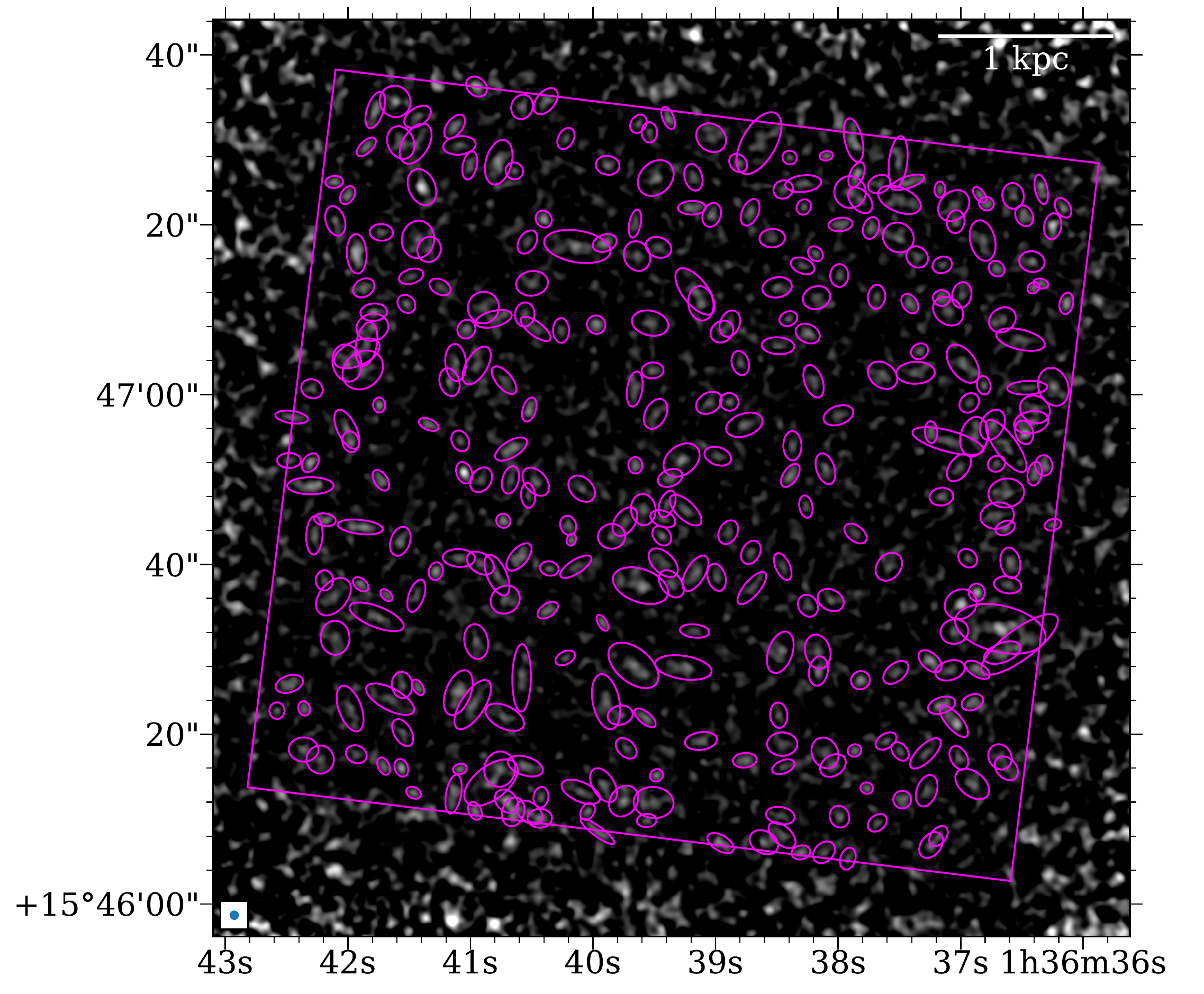}
\caption{The \alma\ band 7 (0.87~mm) continuum map is given on the left. The 80\% footprint is outlined with the box and sources identified using SExtractor are shown. The beam size is given in the lower left corner. The \alma\ band 4 (2.1~mm) continuum map is given on the right with the footprint and sources shown.}
\label{fig:sources1}
\end{figure*}

\begin{figure*}
\plotone{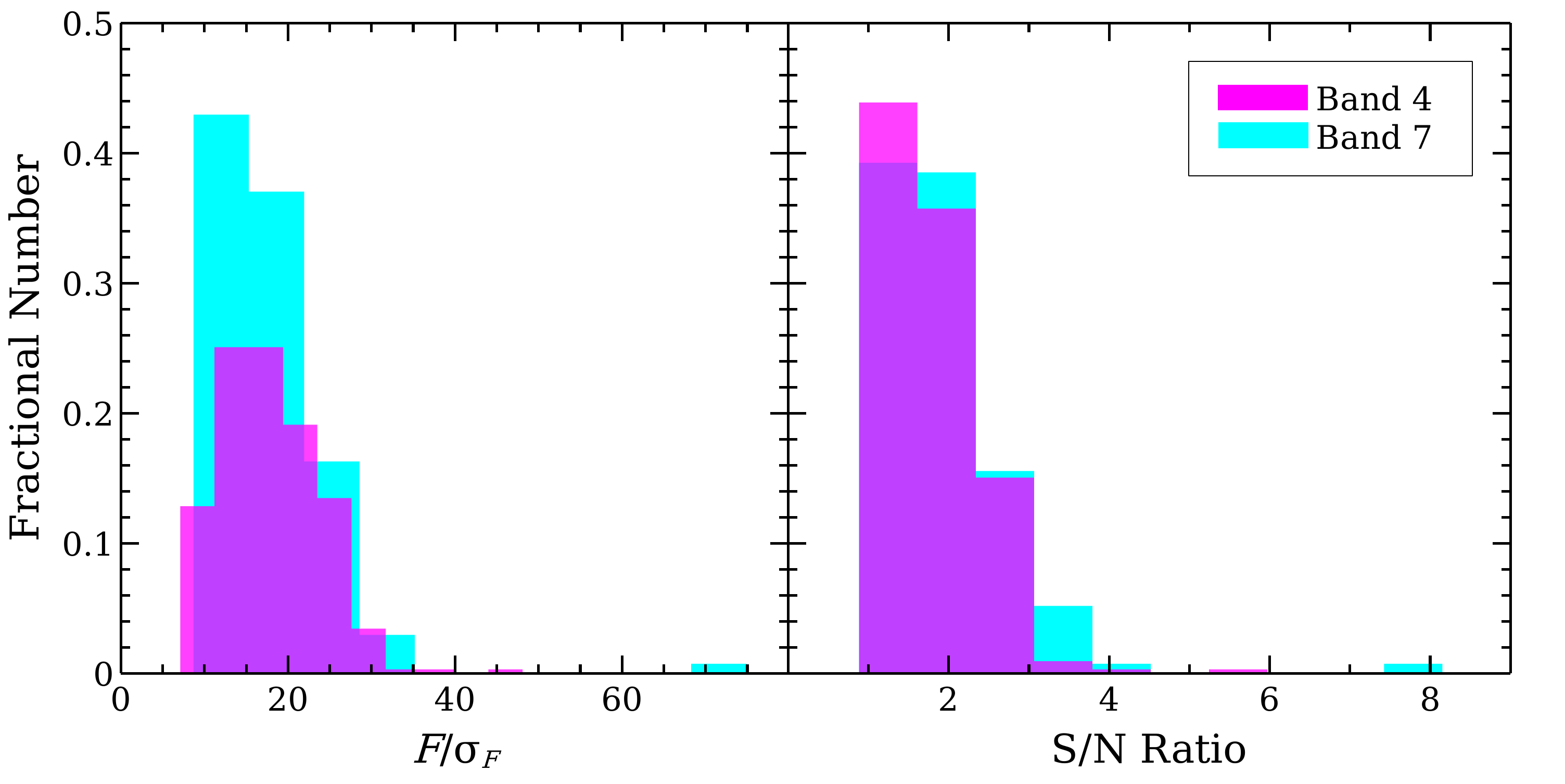}
\caption{Distribution of the flux to the error in flux ratio for each millimeter source found in bands 4 (319 sources, magenta) and 7 (135 sources, cyan). Ratios range from $\sim$\,7 to 35. The brightest source, discussed in Section~\ref{sec:burt}, has a $F/\sigma_{F}$ of $\sim$\,31 in band 4 and 75 in band 7. The distribution of the total signal-to-noise ratio for each millimeter source is given on the left. The S/N for band 4 ranges from $\sim$\,0.6 to 5.4 with a mean of 1.8. For band 7, the S/N ranges from $\sim$\,0.9 to 8.2 with a mean of 1.9. }
\label{fig:snr}
\end{figure*}

\begin{figure*}
\epsscale{0.8}
\plotone{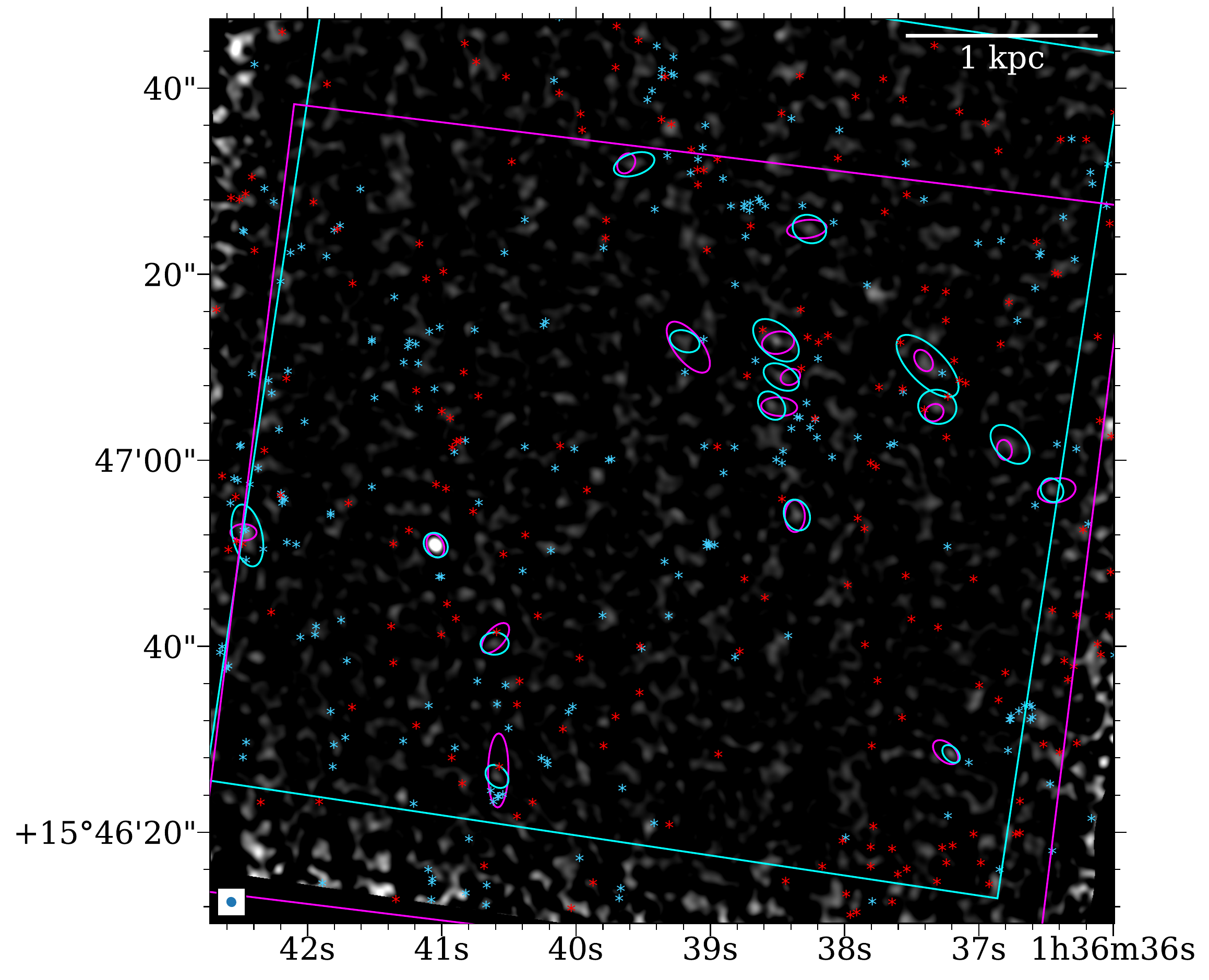}
\caption{The \alma\ band 7 (0.87~mm) map is shown in grayscale with the 16 overlapping sources identified (band 7 ellipses shown in cyan and band 4 ellipses shown in magenta). The 80\% primary beam footprints (band 7 in cyan and band 4 in magenta) are also overlaid to show the 1.8~arcmin$^{2}$ (15.1~kpc$^{2}$) region where the footprints overlap. The position of the LEGUS star clusters are shown as star symbols differentiated according to age. Blue stars represent star clusters 10~Myr and younger. Red stars represent star clusters older than 10~Myr.}
\label{fig:overlap}
\end{figure*}

Millimeter sources for each band are identified using SExtractor \citep{sextractor} with the requirement that each source must have 50 contiguous pixels with flux greater than 2$\sigma$ above the background. With the pixel scale of 0\farcs06 per pixel, 50 pixels is about 420~pc$^2$. Only the sources, highlighted by the SExtractor ellipses, that lie within each image's 80\% footprint are kept as shown in Figure~\ref{fig:sources1}. The photometry on each millimeter source yields flux-to-sigma-flux ratios inside the SExtractor ellipses of greater than 7; Figure~\ref{fig:snr} shows the distribution of this ratio for the sources. A noise for each ellipse is estimated by randomly placing ellipses with the same size and shape as the source found by SExtractor and calculating the standard deviation of the ensemble fluxes for that particular ellipse. A new signal-to-noise ratio is then calculated by dividing a source's flux by this noise estimate; the distribution of which is shown in Figure~\ref{fig:snr}. In combining the source identification results from each band, we restrict the final sample to have source centroids in one band fall within the beam size of sources detected in the other band. This yields a final tally of 16 dual-band sources as shown in Figure~\ref{fig:overlap} with the location and fluxes given in Table~1. Given the density of sources found by SExtractor, there is a possibility of chance alignment between each image. In order to quantify this, we randomly place our same SExtractor identified source apertures across each image, count how many overlap, and repeat this process 10,000 times. We find an average of 15 uncorrelated sources randomly overlap, with a standard deviation of 4 sources.

Figure~\ref{fig:overlap} also shows the position of the LEGUS star clusters in NGC~628. We identify the three LEGUS star clusters that are projected to lie closest to each millimeter source. Since the Gaia mission \citep{gaia2016} provides superior astrometry, we recomputed the projected separations using the Gaia astrometry for the star cluster centroids (i.e., the \hst\ LEGUS images were re-drizzled onto the Gaia reference frame). The improved astrometry resulted in changes of only $\sim$\,0\farcs1 in declination and $\sim$\,0\farcs01 in right ascension, which do not significantly affect the results presented in Section~\ref{sec:results}. Additionally, we use the fluxes obtained from SExtractor to compute the millimeter/submillimeter slope via
\be \label{eq:slope}
S(0.87{\rm mm}/2.1{\rm mm}) = \frac{\Delta \log_{10}{F~(\text{Wm}^{-2})}} {\Delta\log_{10}{\lambda~(\text{m})}} \ .
\ee

\begin{deluxetable*}{cccc}

\tablecaption{Location and Fluxes of the 16 Overlapping Millimeter Sources\label{tab:s}
}

\tablenum{1}

\tablehead{\colhead{RA} & \colhead{DEC} & \colhead{$\log_{10}F$(0.87~\text{mm})} & \colhead{$\log_{10}F$(2.1~\text{mm})} \\ 
\colhead{J2000} & \colhead{J2000} & \colhead{W m$^{-2}$} & \colhead{W m$^{-2}$} } 

\startdata
1:36:36.45 & $+$15:46:56.77 & $-$17.52$\pm$0.06 & $-$18.54$\pm$0.03 \\
1:36:36.76 & $+$15:47:01.69 & $-$17.09$\pm$0.05 & $-$18.98$\pm$0.04 \\
1:36:37.20 & $+$15:46:28.41 & $-$17.50$\pm$0.05 & $-$18.45$\pm$0.03 \\
1:36:37.31 & $+$15:47:05.73 & $-$17.05$\pm$0.05 & $-$18.89$\pm$0.04 \\
1:36:37.38 & $+$15:47:10.12 & $-$16.84$\pm$0.05 & $-$18.59$\pm$0.03 \\
1:36:38.26 & $+$15:47:24.85 & $-$17.30$\pm$0.05 & $-$18.68$\pm$0.04 \\
1:36:38.35 & $+$15:46:54.10 & $-$17.29$\pm$0.05 & $-$18.79$\pm$0.04 \\
1:36:38.47 & $+$15:47:08.93 & $-$17.33$\pm$0.05 & $-$19.06$\pm$0.05 \\
1:36:38.51 & $+$15:47:12.88 & $-$17.02$\pm$0.05 & $-$18.72$\pm$0.04 \\
1:36:38.54 & $+$15:47:05.86 & $-$17.38$\pm$0.05 & $-$18.68$\pm$0.04 \\
1:36:39.19 & $+$15:47:12.78 & $-$17.32$\pm$0.05 & $-$18.36$\pm$0.03 \\
1:36:39.57 & $+$15:47:31.80 & $-$17.29$\pm$0.05 & $-$19.05$\pm$0.05 \\
1:36:40.59 & $+$15:46:25.99 & $-$17.58$\pm$0.06 & $-$18.47$\pm$0.03 \\
1:36:40.60 & $+$15:46:40.27 & $-$17.49$\pm$0.06 & $-$18.45$\pm$0.03 \\
1:36:41.04 & $+$15:46:50.86 & $-$16.69$\pm$0.04 & $-$18.47$\pm$0.03 \\
1:36:42.45 & $+$15:46:51.90 & $-$16.88$\pm$0.05 & $-$19.05$\pm$0.06 \\
		   & Sum		    & $-$15.95$\pm$0.01 & $-$17.45$\pm$0.01
\enddata
\end{deluxetable*}

\section{Results \& Discussion}
\label{sec:results}

\begin{figure*}
\epsscale{0.55}
\plotone{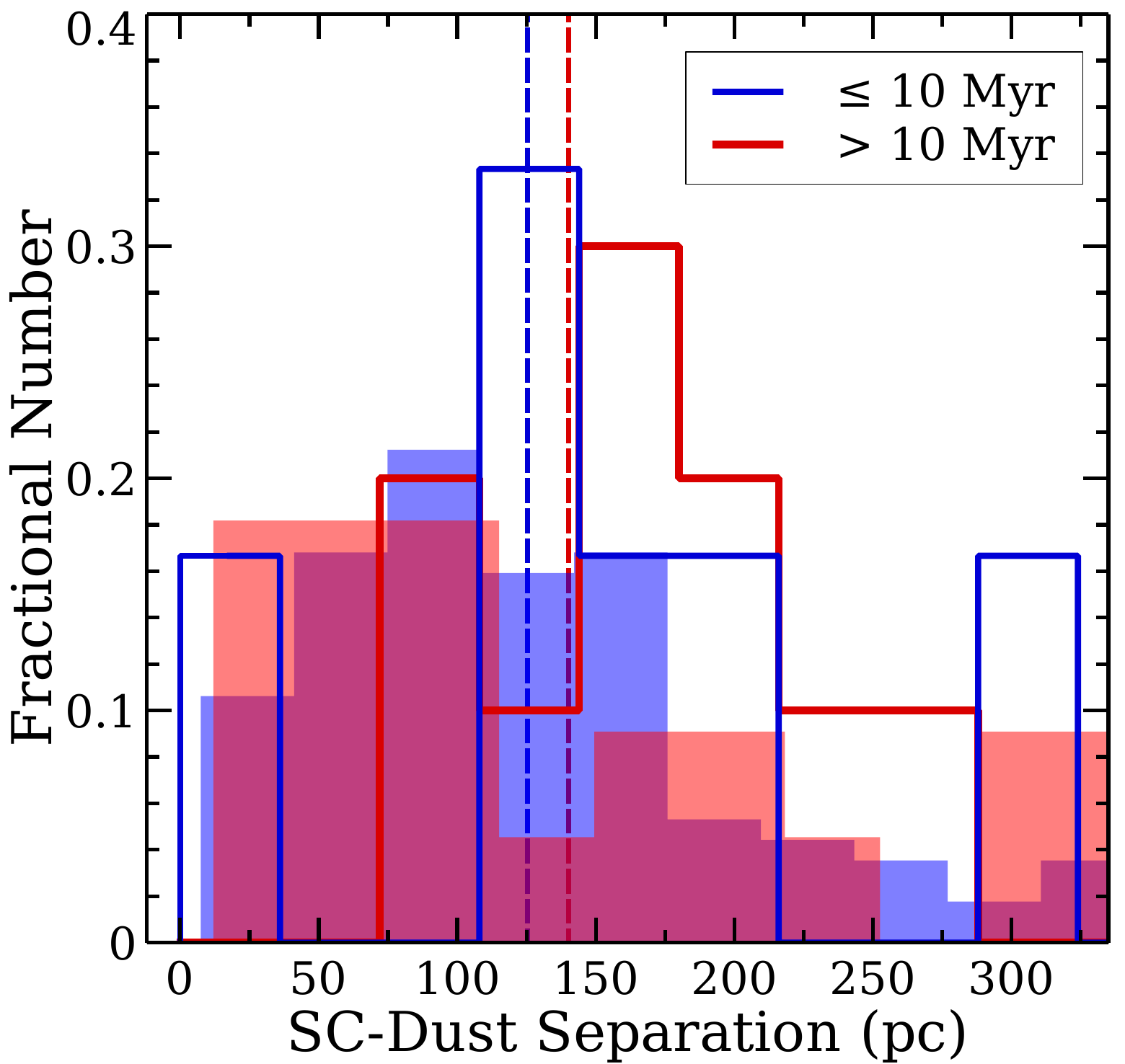}
\caption{Histogram of the physical separations between the 16 overlapping millimeter sources and their nearby star clusters shown by the solid lines. The histogram of separations between all millimeter sources found in band 7 and the nearby star clusters is shown as the shaded regions. Star clusters younger than 10~Myr are given in blue and clusters older than 10~Myr are shown in red. The average star cluster-millimeter source separations, for all band 7 sources, are given as colored dashed lines. The older star clusters, on average, lie slightly further from the dust clouds which is in agreement with the results for star clusters and giant molecular clouds in NGC~5194 from \cite{Grasha2019}.}
\label{fig:hist}
\end{figure*}

Figure~\ref{fig:hist} shows the distribution of the projected physical separations between the 16 overlapping millimeter sources and their associated nearby star clusters. The star clusters are split into two age bins: younger than 10 Myr and older than 10 Myr. The average separation between the 16 millimeter sources and star clusters is $\sim$\,161$\pm$21~pc and $\sim$\,172$\pm$13~pc, respectively, for the younger and older clusters. Since all the sources found in band 7 have high S/N, we include the separations between these sources and their nearby star clusters in Figure~\ref{fig:hist} as the shaded regions. It is possible we detect these sources in band 7 but not in band 4 because their flux falls below the detection limit at the longer wavelength. In order to quantify this, we adopt the average dust continuum slope of $-3.92$ and find 57 of the 135 band 7 sources would be above the band 4 detection limit of 31.7\ujybm. Using the slope of the brightest source (\S\ref{sec:burt}), we find only 9 of the 135 sources would be bright enough to detect in band 4. The new average separation between the millimeter sources and the star clusters is $\sim$\,125$\pm$7~pc and $\sim$\,140$\pm$22~pc, respectively, for the younger and older clusters. This slight difference in projected separations between young and old star clusters, albeit statistically insignificant, still echoes a similar result by \cite{Grasha2019} for the spiral galaxy NGC~5194: the younger LEGUS star clusters have, on average, smaller separations from the nearest giant molecular clouds than do the older star clusters. Again, given the uncertainties in these average separations, there is no statistical difference between the typical projected distance to old and young star clusters.

\begin{figure*}
\epsscale{0.8}
\plotone{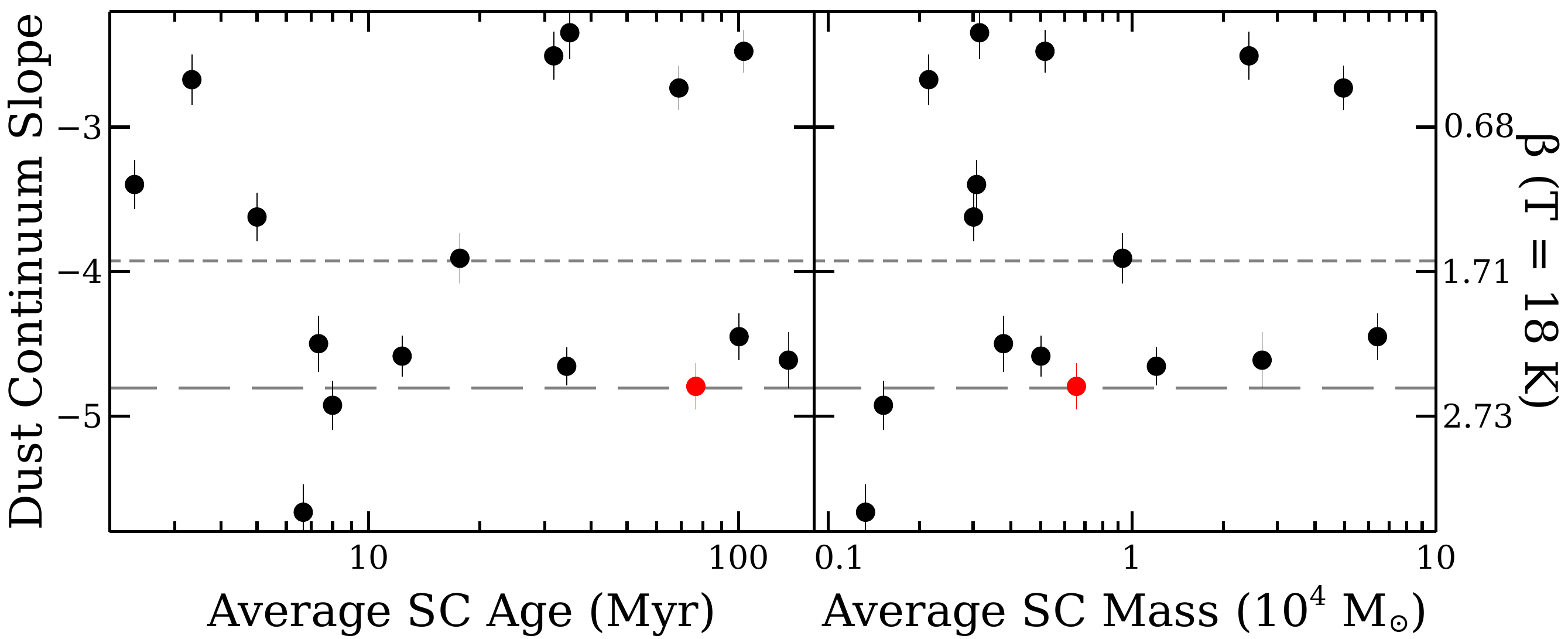}
\caption{The dust continuum slope for each of the 16 sources as calculated in Equation~\ref{eq:slope} plotted as a function of the average star cluster age (left) and mass (right). Shown in red is the brightest source as discussed in Section~\ref{sec:burt}. There is no correlation between the continuum slope and either star cluster age or mass. The long-dashed line designates the slope above which is an excess in dust millimeter emission as determined by the SED given in Figure~\ref{fig:sed}. The short-dashed line marks the slope of the sum of all 16 sources at $-3.92\pm0.3$ which is also shown in Figure~\ref{fig:sed}. See \S\ref{sec:results} for a description of how the uncertainty on the slope was estimated. Also given are the $\beta$ values assuming a temperature of 18 K for a dust continuum slope of $-3$, $-4$, and $-5$.}
\label{fig:sc}
\end{figure*}

Finally, we track the millimeter continuum emission with the nearby star cluster properties. Figure~\ref{fig:sc} gives the 16 sources' millimeter/submillimeter continuum slopes as a function of the nearest three star clusters' ages and masses averaged together. Overplotted in Figure~\ref{fig:sc} is the reference slope $S$(0.87mm/2.1mm)$_{\rm ref}=-4.81$ provided by the dust model SED described in Section~\ref{sec:burt}; slopes larger than this value imply a flatter dust emissivity than the reference dust template. Thirteen of the 16 sources exhibit millimeter/submillimeter slopes larger than this reference value. Also plotted is the slope of the sum of all 16 sources at $S$(0.87mm/2.1mm)$_{\rm sum}=-3.92\pm0.3$ which implies a small but real excess for the sources. The uncertainty on this slope is approximated by the difference between $S$(0.87mm/2.1mm)$_{\rm sum}$ and the slope $S$(0.87mm/2.1mm)$_{\rm stack}$ obtained after stacking all 16 sources with 30\arcsec cutouts and extracting the photometry from the 0.87~mm and 2.1~mm stacks using the same approach outlined in \S\ref{sec:analysis}. One caveat to this excess measurement---recent laboratory measurements of dust grain emissivity show certain dust compositions can lead to a flatter millimeter/submillimeter slope than is expected by modified blackbody models \citep[see e.g.][]{Demyk2017a,Demyk2017b}.

We find no correlation with either star cluster mass or star cluster age, the latter being a proxy for the hardness and/or intensity of the local ISRF.  This null result implies the radiation fields produced by the nearest star clusters do not directly affect the shape of the millimeter/submillimeter continuum through the processing of dust grains and their emissivity properties.  However, two provisos should be clarified.  First, it is possible that more sensitive millimeter/submillimeter continuum observations that detect many more sources could yield different results. This is possibly due to the dust associated with the star clusters having already been dispersed by the time we can observe the clusters at optical wavelengths.  Second, as noted in the Introduction, millimeter/submillimeter excesses are primarily observed in low metallicity environments.  Our observational footprint only extends to a galactocentric distance of $\sim$\,4~kpc, where the \HII\ region metal abundance for NGC~628 is still approximately solar \citep{Moustakas2010}. 

\subsection{Bright Unknown Source}
\label{sec:burt}

\begin{figure*}
\epsscale{0.55}
\plotone{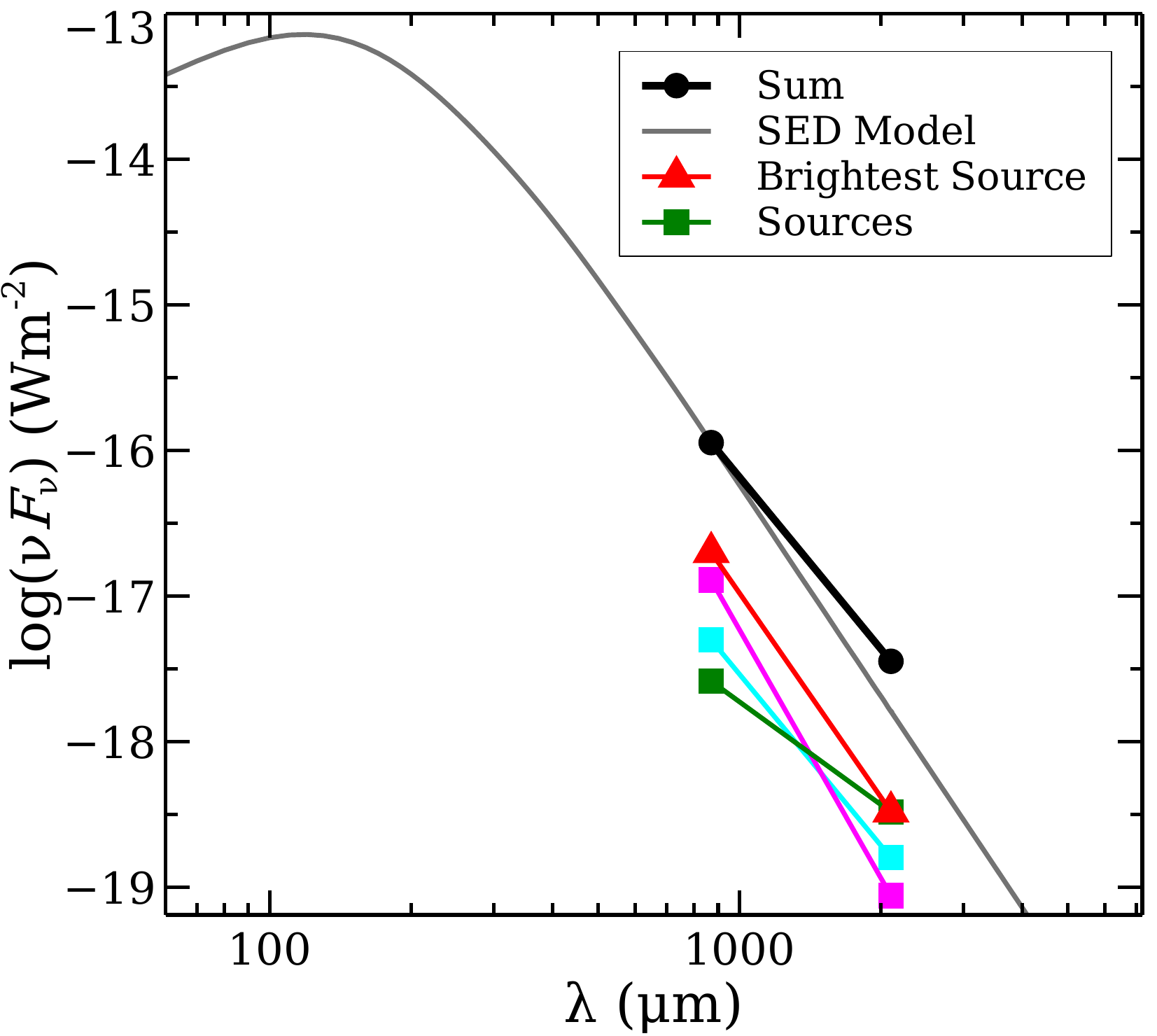}
\caption{The Rayleigh-Jeans tail of a galaxy SED model with cold dust ($\alpha = 2.5$ from \cite{Dale2014}) at a redshift of zero is shown in gray. The colored squares are a selection of three sources that illustrate the full range of dust continuum slopes measured in our sample. The brightest source, as discussed in \S\ref{sec:burt}, is shown as the red triangles. The black circles are the sum of the fluxes from all 16 sources. The model's flux has been shifted to match the sum at 0.87~mm. This particular SED model was chosen because it best fit the slope of the brightest source in our sample. Warmer dust templates from \cite{Dale2014} did not fit as well. The sum shows indicates a slight excess at 2.1~mm. }
\label{fig:sed}
\end{figure*}

The brightest source in our \alma\ maps is found at 1$^{\text{h}}$36$^{\text{m}}$41$^{\text{s}}$.04, $+$15\degr46\arcmin50 86\arcsec (J2000). It is only found in our \alma\ bands 7 and 4 observations; there is no emission found in archival \galex\ ultraviolet (observed 2003), \hst\ optical (observed 2013), \hst\ near-infrared (observed 2005), ground-based H$\alpha$ (observed 2001), \spitzer\ mid-infrared (observed 2004), archival \alma\ bands 6 and 3 continuum maps (observed 2013 and 2015), or CO (2--1) emission line maps (observed 2013). In order to determine if this source is a part of NGC~628 and not a background source like a quasar, we determine the most likely photometric redshift using the observed 0.87~mm and 2.1~mm photometry and the star-forming SED templates of \cite{Dale2014}. The best match is a standard cool dust template SED at redshift zero. The particular template we adopt is the ``$\alpha=2.5$'' model, where $\alpha$ is the exponent in a power-law distribution of different localized infrared SEDs. Fitting a modified blackbody to the ``$\alpha=2.5$'' model yields a blackbody temperature $T=23$ K and $\beta=1.8$. Hence, given the information at hand, this millimeter source is likely within NGC~628. The Rayleigh-Jeans tail of this model dust SED is shown in Figure~\ref{fig:sed} with the the brightest source's flux shown as red triangles. Also shown are three additional example sources to illustrate the range of dust continuum slopes we measure. The photometry for all 16 sources is summed and shown as black circles; the model SED has been scaled to match the flux of the sum at 0.87~mm. For the combined 16 sources, we find a slightly shallower slope compared to the model SED implying a perceptible millimeter/submillimeter excess over what is expected from the cool SED dust template.

We can only speculate on the nature of this peculiar source; it may be a compact and cold infrared dark cloud that is unresolved with our ALMA observations. If this was the case, we would expect to observe the source in the archival \alma\ band 6 ($\sim$1.2~mm, 243~GHz) continuum map taken in 2013 with a comparable beam size of about 1\arcsec and a sensitivity of 850\ujybm. Interpolating the source's flux in band 7 down to the band 6 wavelength gives an expected flux of $\sim$1800\ujybm. The source should have been detected in the band 6 data if it existed at the time of the observation but it is not detected. Therefore, we can rule out the infrared dark cloud explanation. It could alternatively have resulted from a supernova event triggered before the ALMA observations but after the collection of ancillary data mentioned above were taken (follow-up ground-based optical observations are planned). It is possible for this source to be a dust-obscured supernova as it would only take an $A_{V}$ on the order of a few magnitudes to dim the event enough to not be detectable in optical surveys \citep[see e.g.][]{Jencson2017}. Ultimately, deeper imaging at many wavelengths coupled with sensitive spectroscopic information (e.g., \jwst) is needed to more fully understand this enigmatic source. However, if it is indeed a transient event, it may not still be there if observed later with \jwst.

\section{Summary}
\label{sec:sum}

We present results of a joint \hst-\alma\ study of the millimeter/submillimeter emission covering $\sim$15~kpc$^2$ of the nucleus and disk of the nearby spiral galaxy NGC~628 in an attempt to understand how variations in dust emissivity may be affected by local star clusters. We detect 16 sources that spatial overlap in bands~4 and 7 and measure the slope of their dust continua over this 0.87--2.1~mm wavelength baseline. We note, however, that a significant fraction of these spatial overlaps may arise by chance. We find the closest star clusters, as given by LEGUS, to each millimeter source and our analysis suggests that younger star clusters lie closer to the dust than older clusters. This result is in agreement with \cite{Grasha2018} with ALMA-LEGUS observations of NGC~7793 and \cite{Grasha2019} in M51 who find younger star clusters lie closer on average to giant molecular clouds than the older clusters. Tracking the millimeter/submillimeter continuum slopes with the nearby star cluster ages and masses gives no correlation, suggesting that the hardness and/or intensity of local ISRF generated by the star clusters is not responsible for the flatter dust emissivity we observe.  However, we have detected a relatively small number of millimeter/submillimeter sources, and mostly probed the nuclear and central disk regions of NGC~628 where the metallicity is approximately solar and there is a relative dearth of young massive star clusters \citep{Shabani2018}.  It would be interesting to carry out more sensitive continuum observations further out in the disk where the metallicity is lower and any excess dust emission is more likely to appear.

\acknowledgements
We thank the referee for the helpful recommendations. This paper makes use of the following ALMA data: ADS/JAO.ALMA\#2012.1.00650.S, ADS/JAO.\\ALMA\#2013.1.00532.S, ADS/JAO.ALMA\#2016.1.01435.S. ALMA is a partnership of ESO (representing its member states), NSF (USA) and NINS (Japan), together with NRC (Canada), MOST and ASIAA (Taiwan), and KASI (Republic of Korea), in cooperation with the Republic of Chile. The Joint ALMA Observatory is operated by ESO, AUI/NRAO and NAOJ. The National Radio Astronomy Observatory is a facility of the National Science Foundation operated under cooperative agreement by Associated Universities, Inc. We thank Eric Murphy and Laurie Rousseau-Nepton for valuable discussions. A.A. acknowledges the support of the Swedish Research Council, Vetenskapsr{\aa}det, and the Swedish National Space Agency (SNSA). J.A.T. was supported by the Wyoming NASA Space Grant Consortium.  These observations are associated with program \#~13364, the support for which was provided by NASA through a grant from the Space Telescope Science Institute (STScI). Based on observations obtained with the NASA/ESA {\it Hubble Space Telescope}, at STScI, which is operated by the Association of Universities for Research in Astronomy, Inc. under NASA contract NAS 5-26555. This research is partially support by by NSF grants 1716335 (PI: K. Johnson) by the Association of Universities for Research in Astronomy, Inc. under NASA contract NAS 5-26555. 

\facilities{ALMA, HST (ACS, UVIS)}

\software{APLpy \citep{aplpy},
Astropy \citep{astropy1,astropy2}, 
CASA \citep{casa2007}, 
NumPy \citep{numpy},
SExtractor \citep{sextractor}}


\end{document}